# What Makes a Research Article Newsworthy?

Harish Varma Siravuri
Northern Illinois University
DeKalb, IL, USA
hsiravuri@niu.edu

Hamed Alhoori
Northern Illinois University
DeKalb, IL, USA
alhoori@niu.edu

**ABSTRACT**
There has been tremendous growth in the amount of scientific literature being published every year. Yet, very little of it receives press coverage. Mentions by news outlets often establish the relevance the research has to society in general. In the present study, we focused on better understanding the factors that contribute to a research article's newsworthiness. We have built three classifiers to predict the likelihood of research article receiving online press coverage, based on features that quantify the attention it has received on various online platforms. The Random Forest classifier performed best with an accuracy rate of 0.92.

## INTRODUCTION
Research findings are often the subject of news headlines. This is especially true when the topic is of interest to the public or when the findings have a perceptible impact on society. Based on an assessment of the findings as exciting or as particularly relevant to the readers, science journalists present stories about research they think their readers are likely to find interesting. And with the rise of fact-checking journalism (Graves & Glaisyer, 2012), it is generally assumed by the audience that the validity of the findings are confirmed prior to being presented by the news outlets. Consequently, the validity and the relevance of the research are reinforced when mentioned by news outlets. Identifying the factors that go into deciding whether a research article is newsworthy would enable researchers to better position their work to draw attention. To understand what makes a research article newsworthy, we conducted a study to determine whether a relationship exists between the attention an article receives on social media platforms and the attention it receives in news outlets as a basis for predicting the likelihood of newsworthiness.

## RELATED WORK
Fitzpatrick (1999) considered factors that might had an impact on why certain research articles are considered newsworthy whereas others are not considered to be so. These factors include the prestige of the journal, prepublication publicity, and the relevance of the findings to a given audience. Badenschier and Wormer (2012) analyzed the process through which scientific issues are selected for coverage. They concluded that using factors developed for traditional subjects like politics and culture to determine if a story is newsworthy could be misleading and that a certain adaptation was necessary. Rensberger (1978) identified and analyzed three factors that go into making science news: the number of people affected, the trustworthiness of the work, and the fascination value. In earlier studies, researchers criticized inaccurate coverage of published scientific papers (Schwartz, Woloshin, & Welch, 1999), overstatement of results (Lebow, 1999), and sensationalism (Myers, 1996). Allan (2009) identified factual inaccuracies in news reports and how important it is to understand factors that made stories newsworthy in the first place.

## DATA COLLECTION
The data used in the present study were provided by altmetric.com. The database dump consisted of data from more than 5 million articles, which we divided into two categories based on the class label: news. The first category consisted of research articles mentioned in at least one news item, and the second category comprised research articles that had not received any news coverage. We randomly selected 50,000 research articles from each category and extracted information regarding how much attention each research article had received on social media. The outcome was a dataset of 100,000 articles without a class imbalance.

## FEATURE SELECTION
The dataset included a large set of variables pertinent to online attention. Initially, we tried using all the available features. Later, we filtered out certain features based on their sparsity and lack of relevance. The fields that described activity on Pinterest and StackOverflow and the field describing citations in policy were very sparse and could not have contributed much to a research article's newsworthiness. As it was discontinued in March 2013, Connotea was irrelevant to research articles published after that year. We relied on the number of mentions on Twitter, Wikipedia, Google+, Weibo, Facebook, videos, blogs, and peer reviews.

## METHODS

To predict how likely a scholarly publication is to attract news coverage, we built three classification models: a Support Vector Machine (SVM) model using a Radial Basis Function (RBF) kernel, a Random Forest (RF) model with 100 decision trees, and a Multinomial Naïve Bayes (MNB) model. We trained the models using a training set consisting of 80% of the original data. The remaining 20% of the data were used as a test set to evaluate the models. We built all three models using 10-fold cross-validation. We also calculated the weight for each feature to determine its relative importance in the decision-making process. SVM, however, could not be subjected to the same treatment because of the use of the RBF kernel, and feature weights can be calculated only for linear kernels.

## RESULTS

The classification models performed well, with the Random Forest classifier delivering the highest accuracy rate of 0.924. The accuracy, precision, and recall values and the F-1 scores of each of the models are presented in Table 1. The receiver operating characteristic (ROC) curves and the area under the ROC curves are shown in Figure 1.

|  | MNB | RF | SVM |
|---|---|---|---|
| **Accuracy** | 0.782 | 0.924 | 0.888 |
| **Precision** | 0.302 | 0.796 | 0.806 |
| **Recall** | 0.365 | 0.658 | 0.326 |
| **F1-Measure** | 0.331 | 0.720 | 0.465 |

**Table 1.** Accuracy, precision, recall and F1 scores

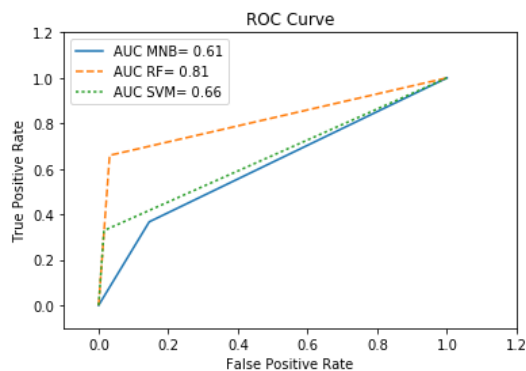

**Figure 1.** ROC curves for each classification model

The five most significant features in respect to the RF and the relative importance of each in respect to the MNB classifier are shown in Table 2. Counts on Mendeley proved to be most significant to the RF model whereas the Video feature proved to be the least significant. The opposite was true in case of the MNB model.

| Feature | Random Forest | MNB |
|---|---|---|
| Mendeley | 0.168083 | 0.5792 |
| Facebook | 0.151553 | 2.8116 |
| Twitter | 0.147885 | 1.3097 |
| Blogs | 0.106562 | 4.0367 |
| Google+ | 0.093940 | 3.5126 |

**Table 2.** Relative significance of features

## CONCLUSIONS AND FUTURE WORK

In this study, we used features that describe the attention research articles receive online to build classification models that predict whether an article is likely to receive news coverage. The RF model delivered good results that imply the existence of a relationship between the attention a research article received online and the likelihood of it receiving news coverage. The results are in agreement with the growing opinion that the newsworthiness of research is increasingly being determined by the readers who post content about it online. In future research, we plan to improve the classification models and build regression models to predict the number of mentions a research article is likely to receive from news outlets.